\documentclass[a4paper,12pt]{article}
\usepackage[cp1251]{inputenc}
\usepackage{graphics}
\usepackage{amssymb,amsmath}
\usepackage[english]{babel}
\usepackage{indentfirst}
\begin{document}
\bigskip
\centerline {\bf ON LAMINAR CONVECTION IN SOLAR TYPE STARS}

\bigskip

\centerline {E.A.Bruevich $^{a}$, I.K.Rozgacheva $^{b}$}

\bigskip

\centerline {Sternberg Astronomical Institute, Moscow, Russia}\
\centerline  {Moscow State Pedagogical University, Russia}\

\centerline {E-mail: $^a${red-field@yandex.ru},
 $^b${rozgacheva@yandex.ru}}\

\bigskip
       We present a new model of large-scale multilayer convection
       in solar type stars.
       This model allows us to understand such self-similar
       structures observed at solar surface as granulation,
       supergranulation and giant cells. We study the slow-rotated
       hydrogen star without magnetic field with the
       spherically-symmetric convective zone. The photon's flux
       comes to the convective zone from the central thermonuclear
       zone of the star.
       The interaction of these photons with the fully ionized
       hydrogen plasma with $T>10^5K$ is carried out by the
       Tomson scattering of photon flux on protons and electrons.
         Under these conditions plasma is optically thick relative
         to the Tomson scattering.
       This fact is the fundamental one for the multilayer convection
       formation.
       We find the stationary solution of the convective zone
       structure.
       This solution describes the convective layers responsible
       to the formation of the structures on the star's surface.

\bigskip

KEY WORDS Large-scale convection, Tomson scattering, solar
atmosphere structures

\vskip12pt
 INTRODUCTION
\vskip12pt
           The systematic extreme ultraviolet and X-ray emission
       observations from Skylab station, Yohkoh, SoHO and Trace satellites
       give us the very interesting images of solar corona.
       After the previous images modifications (the partial gain
       of some interesting details) we can see large-scale corona
       structures around the solar disk and these structures are not
       associated with active regions (Priest et al., 1990; Chertoc, 2002).
       The structures are
       similar to standard coronal loops that connected separate
       active regions together (Beck, 1998), but their "foots" lean on
       the photosphere out of active regions. These regular structures
       cover the hole solar disk as the more large-scale chromocpheric
       network.
       The lifetime of such loops is about a week for relatively
       small loops with length approximately equal to
       $2 \cdot 10^4$ $km$ and with plasma concentration approximately
       equal to $10^{15}$ $m^{-3}$ but for the most great loops
       with length approximately equal to $3 \cdot 10^5$ $km$
       and with plasma concentration approximately
       $2 \cdot 10^{14}$ $m^{-3}$ the lifetime is about some months.
         These large-scale structures (chains, loops) are observed
       for some years. We see that all the observed loop systems
       associated with quiet Sun permanent exist as a regular part
       of solar corona.
          It's necessary note that photosphere and chromosphere
       have regular structures such as grains, supergrains
       and giant grains. The giant grains are discovered by the
       helioseismology's methods (Bec, 1998). These giant grains have the
       regular structure, their sizes are about $3 \cdot 10^5 km$
       with regular plasma speeds of $100 m/s$.
       It's well known that granulation, chromospheric network,
       supergranulation and giant loops is the consequence of
       under-photosphere convective zone existence.
       The solar-like stars photospheres have the similar
       structures as Sun has: grains and supergrains.

       In this paper we present the simple model of the hydrogen
       star convection zone.

       The necessary condition of free convection (rises in plasma
       layers with thickness of only some times smaller then solar radius)
       is the Schwarzschild criterion -- the specific entropy of
       plasma decreases with moving away from the star center.
       Such convection will develop when the temperature inside of
       small convective volume (convective cell) decreases slower
       than the temperature decreases in neighboring plasma (Rudiger, 1989).

       We use the Schwarzschild criterion later in our paper.

       If solar convection is laminar so such processes as
       granulation, chromosphere network and supergranulation
       may exist in the convective layers of different
       thickness. Therefore solar convective zone consists of
       the three layers at least.

       Under large-scale laminar convection conditions the
       small-scale turbulent convection appears owing to
       development of different plasma instabilities.
       The ejections of matter at granulation and supergranulation
       scales are connected more than likely with instabilities
       dynamics.
       The lifetime of these ejections is small and they can't
       change the regular structure of quiet Sun but these
       ejections outline this structure by effective way.

       In this paper we propose the laminar convection model.
       In this model all the structures --
       granulation, chromosphere network, supergranulation
       and regular large-scale coronal structure -- are
       examined as result of laminar convection action. We
       suppose that these structures are the different
       realizations of solution spectrum.

       Let's consider the main assumptions of this model:
       The convective zone is the layer with the spherically symmetry
       distribution of plasma around the radiative transfer energy zone.
       In this layer the condition of real hydrostatic
       equilibrium is carried out.

       In this paper we consider a case when the layer
       under study consists of the ionized hydrogen plasma only
       (protons and electrons).
       This consideration may not be applied for the Sun
       atmosphere conditions but it significantly simplifies the
       mathematical description of convection and allow us understand
       the mechanism of convection zone structures formation.
       In our model this layer is open system through which
       the energy flux moves upwards. So let's consider
       that the plasma conditions we can described as
       polytropic equation:

              $$  {N\over N_0} =
        \Bigg( {T\over T_0}\Bigg)^{n} \eqno (1) $$

       where n is the polytropic index, N and T are the
       plasma concentration and temperature. These values
       are $N_0 \approx 5\cdot 10^{27} m^{-3}$ and
       $T_0 \approx 2\cdot 10^{6}K$ at the bottom border
       of the layer.
       The layer thickness (the convective zone depth) is
       approximately 0.3$R_{\odot}$, where $R_{\odot}$ is the
       solar radius.
       The energy emission come to convective zone, the
       temperature  ${T_r} \approx T_0$ near the bottom border
       of the convective zone.
       This flux is the reason of the development of laminar
       convection.

       The emission and ionized plasma interaction is carried out
       by photon scattering on electrons and protons
       in case where the photon energy don't exceed the
       value $kT$. The time of energy transmission
       from photons to plasma don't excess the value (Kaplan, Tsytovich, 1973):
              $$  {t_0} = {{3m_pc}\over 8\sigma_T\epsilon_r}, \eqno (2) $$
       if $kT_r\ll m_pc^2$, where $k$ is the Boltzmann's constant,
       $\sigma_T$ is the Thomson probability section of scattering,
       $m_p$ is the electron mass,
       ${\epsilon_r} = {{{4 \sigma_B} \over {c}} {T_r}^4 }$ ,
       $\sigma_B$ is the Stephan -- Boltzmann constant.
       If $T_r = T_0$ then $t_0 \approx 0.1 s$.

       The distance of free run for photons is equal to
       $ \Delta \approx (\sigma_T N)^{-1}$. If $N = N_0$ then
       $\Delta \approx 3 $ m.

       In $\Delta^3$ volume plasma and emission are in
       thermodynamic equilibrium almost because of the radiation
        is connected with matter.

       The thermal conductivity mechanism is made available by
       the next processes in our case. The plasma (heated by
       radiation in value $\Delta^3$) loses the energy by
       bremsstrahlung. The speed of these losses is
        $\epsilon_{ep} = 1.6 \cdot
       10^{-40}N^2 \sqrt{T}$  $J\cdot m^{-3} s^{-1}$.
       The characteristic time of this process is equal to
       (Kaplan, Tsytovich, 1973): $t_2 = 2.6 \cdot 10^{17} \sqrt{T} N^{-1} s$.
       If $N=N_0, T=T_0$ we derive $t_2\approx10^{-6} s$.
       At the other hand the bremsstrahlung
       heats up the electrons in the vicinity of the volume
       $\Delta^3$. The time taken for this process
        $$  {t_1} = {{3m_ec}\over 8\sigma_T E}, \eqno  (3) $$
        where $E = 3/2 N k T m_e/m_p$ is the electron
        energy density.
        If $T_* \le T \le T_0 $ $and$ $N_* \le N \le N_0, N_*=2.5 \cdot
        10^{26} m^{-3}$ we find $20s<t_1<160s$.

        The characteristic rate of thermal
        conductivity is equal to ${v_{\chi}}= {{\Delta}\over{t_1}}$.
        So the thermal conductivity coefficient for the process
        described (for the same order of magnitude) is equal to:
        $$ {\chi} = {v_{\chi}\lambda} ={{\lambda \Delta}\over {t_1}} \eqno (4),$$
        where $\lambda$ is the thickness of the shell warmed up.

        The convectional energy transfer is carried out thanks to
        macroscopic transports of the value $\Delta^3$.
        The temperature inside the volume $\Delta^3$ is higher then
        the plasma temperature in the layers which are situated higher
        then the bottom border of the convective zone, see Section 1.
        Thus the Archimedean raising force acts on this value
        and gives him the acceleration $g\cdot {\Delta T}\over{T}$,
        where $g$ is the free fall acceleration on the
        bottom border of the convective zone, ${\Delta T} =
        {T_0 - T}, T<T_0.$

        The flotation process is retarded by viscosity.
        In our case
        the viscosity is the consequence of the Tomson
        scattering. The value $\Delta^3$ is full of plasma and radiation.
        When this value moves the radiation is scattered
        by electrons of neighboring plasma.
        Thanks to the scattering the equalization of electron
        momentum
        takes place inside the volume $\Delta^3$ and outside of one.
        This viscosity they called radiation viscosity.
        It characterized by the viscosity coefficient
         $$ {\nu} = {{1}\over {3}}{{c}\over {\sigma_T N}} \eqno (5).$$

        If $N = N_0$ then ${\nu} \approx {6\cdot 10^9} $ $m^2/s.$
        This value is similar to the value estimation taken from
        the analysis of observations.
        The floating is ended when the raising force is in
        equilibrium with viscosity forces. The characteristic
        time of convective floating is equal to
        $$  {t_2} = {{\nu}\over{g\lambda
         {{\Delta T}\over{T}}}}, $$
        where $\lambda$  is correspond to characteristic scale
        of the convective layer (the mixing length).
        If $\lambda \approx 2\cdot 10^8 m$,
        $T = T_0$ and ${{\Delta T}\over{T}} \approx 1$,
       $g = 2g_\odot$, where  $g_\odot \approx 274 m/s^2$
        is the gravity force acceleration on the solar surface
        then we have
        $t_2\approx 0.05 s$.
        So at the bottom border of the convective zone
        the relation $t_1>t_2$ is taken place. In this
        case the convective transfer is more effective then
        the heat conduction.

        Near the top border of the convective zone
         $N = 4 \cdot 10^{22}$
        $m^{-3}$ and $\Delta_* = ( \sigma_T N)^{-1} \approx 10^3$ $km$.
        In this case we can ignore the Tomson effect.
        The radiation of plasma propagates free up to solar photosphere.

        In the section 1. we give the solutions of stationary convective
        zone structures
        in the hydrodynamics approximation
        with the heat conduction (4) and viscous (5) coefficients.

        These solutions have the solitary wave structure and
        describe the model of multi-layer convection.
        All the convective cells have the torus contour.

\bigskip

1 THE EQUATIONS OF THE STATIONARY CONVECTIVE ZONE STRUCTURE

\bigskip

        The set of simultaneous equations for the spherically
        symmetric stationary convective zone which rotates about $z$ axis
        (because of the hydrodynamics approximation is correct)
        have the form:

        $$ { ( \vec{v}, \nabla) \vec{v}}
        = {{{ \nabla (p + p_r)} \over {\rho}} + \vec {g}
        + \nu \cdot \nabla \vec{v} - 2[ \vec{v}, \vec{\omega}]} \eqno (6) $$

        is the motion equation,
        where $ \vec{\omega}$ is the angular velocity of convective
        zone rotation, $\rho$ is the plasma density,
        $ p$ is the plasma pressure, $ p_r$ is the pressure of radiation.

        $$ {( \vec{v}, \nabla) T}
        = {\chi \cdot \Delta T} \eqno (7) $$

        is the heat conduction equation,

        $$ { \nabla (p + p_r) + \rho \vec{g} } = 0  \eqno (8)$$

        is the hydrostatics equilibrium equation,

         $$ {{ dp_r} \over {dr}} =
         {- {{ \sigma_T N} \over{c}}{{1} \over {4 \pi r^{2}}} L} \eqno (9)$$

        is the radiation transfer equation outside of the volume $ \Delta^3$,

         $$ {{ d L} \over {dr}} =
          {4 \pi r^{2} \epsilon_{ep}}  \eqno (10)$$

        is the bremsstrahlung of plasma equation inside the volume $ \Delta^3$,

         $$ {dM} =  {4 \pi r^{2} \rho \cdot d r}  \eqno (11) $$

        is the mass conservation equation.

         In the equation (8) we don't take the density of radiation
         $ \rho_r$ because of $ \rho_r \ll \rho$ in solar-like stars.

         The set of simultaneous equations (8-10) have used
        by A.S.Eddington in 1926 (Eddington, 1926).

        Let $p = NkT, N=N(r), T=T(r)$ and state of plasma is described by
        the polytropic equation (1).
        Let's take the variable $x = {{r}\over {\zeta}}$,
        where $$ \zeta =
        {\Big({{kT_0} \over {4 \pi G m_p \rho_0}} \Big)^{1/2}}
        \approx  13 \cdot 10^5 km  > R_{\odot}.$$
        Then function $\tau = {{T} \over {T_0}}$ with equations (8) - (11)
        can be transformed to the following equation
        $$ {(n+1){{1} \over{x^2}}(x^2 \tau_x)_x} =
        - \tau^n + \alpha \tau^{2n+ {{1} \over {2}}},  \eqno (12)$$
        where index $x$ means the differentiation with respect
        to $x$
        and $$ \alpha = {{ \sigma_T \epsilon_{ep}(N_0,T_0)} \over {4 \pi
        G m_p \rho_0}} \approx 10^{17}.$$

        Let velocity vector $\vec{v}$ has the
        $\{ V,W,Z\}$ components in spherical coordinate system. The
        vector of angular velocity $ \vec{ \omega}$ has the following
        components:

        $\{ \omega cos \theta, -\omega sin \theta,0, 0 \}$.

        Let $\theta \ll 1$. From the equations of the structure
       $${(\vec{v}, \nabla) \vec{v}} = {\nu \cdot \Delta \vec{v} -
       2[ \vec{v}, \vec{ \omega}]} $$
       $$ {(\vec{v}, \nabla) T} = { \chi \cdot \Delta T}, \eqno (13)$$
       one can find the equation for $V(x)$ and $\tau(x)$:
       $$ {VV_x} =
    {{ {\nu} \over {\zeta }} \cdot {{1} \over {x^2}} \cdot(x^2 V_x)_x}$$
       $$ {V \tau_x} =
    {{ {\chi} \over {\zeta }} \cdot {{1} \over {x^2}} \cdot(x^2 \tau_x)_x}
        \eqno (14)$$
       The equations (12) and (14) are simplified when we assume
       that the component of velocity $V$ is decreased with the
       depth. This condition is in agreement with solar
       observations: the plasma spread out velocity in the
       photosphere decreases with the scale increasing from grains
       to giant grains.

       We choose the solution in the next form:
       $$ V = {{{\sigma} \over {\zeta}} \nu} \eqno (15) $$
       where $\sigma$ is the free parameter.

       This permits us to simplify the equation
       (14) and transform it to the following:

   $$(n + 1) \beta \tau_x = - \tau^{2n+1}(1 - \alpha \tau^{n+1/2}), \eqno (16)$$
       where $\beta = 7.5, \sigma= {{\nu_0} \over {\chi_0}} \sigma$.
        The equation (16) has different solutions
       for the different values of $n$. Let's choose the
       value $n$ (use the Schwarzschild criterion). According to this
       criterion the temperature
       inside the small element $\Delta^3$  has to decrease with
       increasing of the distance from the star center slower then
       decreasing of plasma temperature occurs. The plasma is
       in the hydrostatic equilibrium and the radiation is absence.

       Substitute $p_r = 0,
       p = NkT$ and $\rho = m_p \cdot N$ to (8).

       Use the polytropic equation (1), we find the relative
       change of the plasma temperature (radiation doesn't take
       into account)
       $$|{{T_x} \over {T}} |_0
       = {{m_pg \zeta } \over {(n+1)kT_0}} \tau^{-1}.$$

       In the volume $\Delta^3$ the temperature changes according to (16).
       Also let's take into account $\alpha \gg 1$ and $\tau \leq 1.$
       Then $\alpha \tau^{n+1/2}\gg1$
       and relative change of the temperature inside the volume
       $\Delta^3$ is approximately equal to:
       $$|{{T_x} \over {T}}| \approx {{\alpha} \over {(n+1) \beta}} \tau^
       {3n+1/2}   $$
       In our case of the evolution of the convective instability
       $|{{T_x} \over {T}} |_0 > |{{T_x} \over {T}} | $ the number $ \beta$ is
       evaluate as:  $ \beta > {{\alpha } \over {223}} \tau^{3n+3/2}.$
         In this case we have $\tau^n \gg (\alpha \sqrt{\tau})^{-1}.$
       Therefore $\beta > (223 \alpha^2)^{-1}.$

       At the other hand one can integrate the equation (16)
       because of ignoring the first member of the right part of the equation.

       Thus we obtain the next algebraic equation:
       $$ \tau^{-3n-1/2} - 1 = {{\alpha} \over {\beta}}
       {{3n+1/2} \over {n+1}}(x - x_0).  $$
       Using this equation and the consideration that
       $ \tau^n \gg (\alpha \sqrt{\tau})^{-1} $
       we find that $\beta < {{3n+1/2} \over {n+1}} {{x-x_0} \over {\alpha^2}
       \tau}.$
       The value of polytropic index $n$ (1) is necessary to
       choose as to make up the next unequality:
       $$  {{1} \over {223 \alpha^2}} < \beta
       < {{3n+1/2} \over {n+1}} {{x-x_0} \over {\alpha^2 \tau}} . \eqno (17)$$
       For the solar-like star we have
       ${{1} \over {303}} \leq \tau \leq 1$
       and $10^{-4} \leq x-x_0 \leq {{2} \over {57}}. $
       In this case the unequality (17) is realized for
       all $n$ having the positive values.
       Let's choose $n = 3/2$. Then the accurate solution of the equation (16)
       for $\tau(x)$  one can find from the next algebraic equation:
       $$ {{{2}\over {5 \beta}}(x - x_0)} = {{{1} \over {3}}(1- {{1} \over
       {\tau^3}}) + \alpha (1 - {{1} \over {\tau}}) -
       {{\alpha^{3/2} \over {2}} \Big({ln {{\alpha^{1/2}+1}
       \over {{\alpha^{1/2}-1}}}} - ln {{\alpha^{1/2} \tau+1} \over
       {{\alpha^{1/2} \tau - 1}}} \Big)}} \eqno (18) $$

       We have taken into account that $\tau(x_0) = 1$ here.

       The solution for $\tau(x)$ has the solitary wave form.
       It's clear from the form of the equation (16).

       If $ \alpha^{1/2} \tau >>1$ we have the asymptotic solution
       $$ {{\Big({{T_0} \over{T}}\Big)}^5} \approx {{{2\alpha} \over
       {\beta \zeta}}(r-r_0)}  \eqno (19)                  $$

       For $x \rightarrow x_0$ one can find that
   $$\tau \approx e^{-{{2} \over {3 \alpha \beta}}(x - x_0)} \rightarrow 1.$$

       At last we can find the expression for the speed components
       $W$ and $Z$.
       Then we examine the most simple case of the symmetric spreading out
       on the sphere surface when $W =Z$. Let's consider also that the
       angular velocity $w$ we can take from the equation
    $$V{{{\partial W} \over {\partial x}} + 2W \cdot \omega \cdot cos \theta} =
      {{{\nu} \over {\zeta^2}} \cdot {{1} \over {x^2}} \cdot
       {{\partial} \over {\partial x}}\Big (x^2
       {{\partial W} \over {\partial x}}\Big )}. \eqno (20)$$
       As follows from the equation (20) the convective zone
       rotates differently.
       Thanks to the convection the redistribution of the rotatory
       moment inside the star takes place. This effect is accurately
       studied in (Rudiger, 1989).

       Under conditions selected in our paper we can find the equation
       for $W$ from the first equation of the set of simultaneous
       equations (14). His form becomes simple enough
      $$ {{d W^2} \over {d l}} =
      {{{\nu} \over {\zeta^2}} \cdot {{1} \over {x}} \cdot
       {{d W^2} \over {d l^2}}} , \eqno (21)$$
       if we change the $\theta$ and $\phi$ angular variables to
       $l$ variable and $dl = \sqrt{d \theta^2 + sin^2 \theta
       d \varphi^2}. $
       Among the multiple numbers of solutions of the equation (21)
       there is periodic solution. This periodic solution
       has the next form:
        $$ W = {W_0 \cdot tg(W_0 {{\zeta^2} \over {\nu}}
        \cdot x \cdot (l - l_0))}, \eqno (22) $$
        where $W_0$ is the speed peak value $W$, the point $l_0$
        is situated at the radius $x$ and is the start
        reading for $l$ coordinate.
        On the surface of sphere with radius $x_*$ plasma spreads
        out from $l_0$ point. So our model is symmetric there are
        many points $l_{0,i}$ on the surface of the sphere of
        radius $x_*$. The distance between the neighboring points
        is equal to $2 \xi = x_0(l_{0,i}-l_{0,i-1})=x_* \cdot \Delta l
        = {{\pi \nu} \over {W_0 \zeta^2}}.$ Between these points
        there are two opposing plasma streams with velocities
        of opposite direction. These streams compensate each other
        at the distance $\xi$ from the each points.

        So all the surface of the radius $x_*$ breaks-down to the
        cells with diameters which are equal to $\xi$.
        All the number of these cells $L$ we can calculate when we the
        surface square $\pi x_*^2$ divide by the cell square
        $\pi (\xi/2)^2$ : $L= 4 \cdot (x_*/\xi)^2.$ Then the
        velocity amplitude is equal to $W_0 = \pi \nu / 2 \xi^2 x_*
        \sqrt{L}.$ The kinetic energy density $\epsilon$ is
        proportional to $W_0^2.$ So the convective streams have
        the spectral energy distribution $\epsilon \sim L \sim
        \epsilon^{-2}.$ The solutions of the convective zone
        structures (21) and (22) describes the stationary
        convection when all zone of the convective energy
        transfer consists of the layers with the different
        thickness. Every convective cell have the torus form.
        These solutions of this important problem are made for the
        first time. From the equation (19) follows the next conclusion:
        the convective zone differently rotates. Thanks to the
        convection the rotation moment redistribution inside the
        star is taken place. This effect is studied in detail in
        (Rudiger, 1989).

\bigskip

2 SUMMARUY AND CONCLUSIONS

\begin{figure}[h!]
 \centerline{\includegraphics{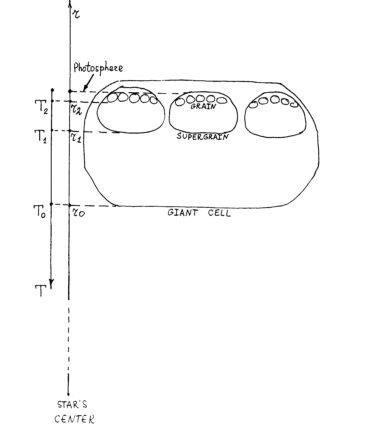}}
 \caption{The convective zone structure}\label{Fi:Fig1}
\end{figure}

\bigskip

         This model qualitatively describes the deep convective layers
         of the star under the supergrains layer. In case of the star's convective
         transfer it's important that plasma at these layers is the fully
         ionized. We don't study star's plasma at the highest
         convective under-photospheric layer where the turbulent
         processes are possible. In this turbulent layer
         there are necessary conditions for the generation of
         the long-scale magnetic field of the star. At the layers
         under this turbulent under-photospheric layer the
         convection is the stationary convection.

         Let's use the asymptotic solution (19) for the convective
         zone analysis. We have the convective zone
         consists of some layers with thickness of $\lambda_i, i=0,1,2,...$
         The temperature on the lower part of the layer's border is
         equal to $T_i$, on the top part is $T_{i^*}$. If we take the
         dependence of parameters from (15) and (19) on $T_0$ into account
         so we can find the relation between the velocity and
         temperature at the bottom and top borders of the
         neighboring layers:
         $$ V_{i-1}/V_i=(\lambda_{i-1}/\lambda_i)^2(T_i/T_{i-1})^{3/2}(T_{i-1^*}/T_{i^*})^5 \eqno(23)$$

         For the qualitative estimation let's substitude the characteristics
         of the convective layers associated with giant cells and supergrains
         into (23):
         $$ V_0=10 m/s, T_0 \sim 2\cdot10^6K, \lambda_0 \sim 3\cdot10^5 km$$
         $$ V_1=100m/s, T_1 \sim 10^6K, \lambda_1 \sim 3\cdot10^4 km.$$
         In this case we obtain that $T_{0^*}\approx0.4T_{1^*}$
         and the temperature on the top border of the layer
         $\lambda_0$ is smaller than the temperature on the top border of
         the layer $\lambda_1<\lambda_0$. So we can see that $\lambda_1$
         torus  are situated into $\lambda_0$ torus.

         This qualitative analysis of the formulae (23)
         allows us to make the conclusion about relatively
         disposed convective layers in the hydrogen star. The
         layers are put one into another as we can see at the
         Figure 1.

        In (Rozgacheva et al., 2003; Rozgacheva et al., 2004 )
        was shown, that the torus typical
        scales may form the geometric progression. This fact is one
        of our model tests. Such geometric progression that
        described stationary structures at Solar surface is probably
        form the fractal set.

         Acknowledgements. The authors thank the RFBR grant 09-02-01010 for support
         of the work.

\bigskip

References

\bigskip

        Beck I.G., Duvall T.L. et al.//Nature. 1998. V.394.
        N6694. P.653.

        Chertoc I.M.//htpp://helios.izmiran.troitsk.ru/lars/Chertok/. 2002.

        Eddington A.S.//The internal constitution of the stars.
        Cambridge. 1926.

        Kaplan S.A., Tsytovich V.N.//Plasma Astrophysics.
        Oxford.:Pergamon-Press., 1973.

        Priest E.R., Foley C.R. et al.//Nature. 1998. V.393.
        N6685. P.545.

        Rozgacheva I.K., Bruevich E.A.//In collected articles "Synergy. The workshop
        proceeding", 2003, V.5, Publishing house of Moscow State
        University, P.199.

        Rozgacheva I.K., Bruevich E.A.//Proc. conf.:"The fractals and their application
        in science and technics", 2004, Tyumen, Russia, P.29.

        Rudiger C.//Differential Rotation and Stellar
        Convection of Sun and Solar-type Stars.
        Berlin:Akademie-Verlag. 1989.

\bigskip

\end{document}